\documentstyle[aps,preprint,epsfig]{revtex}
\tightenlines
\def\h{\leftrightarrow}
\def\s{\updownarrow}
\begin{document}
\title{Tomographic test of Bell's inequality}
\author{G. M. D'Ariano, L. Maccone and M. F. Sacchi}
\address{Theoretical Quantum Optics Group\\
Universit\`a degli Studi di Pavia, INFM --- Unit\`a di Pavia, 
via A. Bassi 6, I-27100 Pavia, Italy}
\author{A. Garuccio} \address{Dipartimento Interateneo di Fisica,
INFN, Sezione D, I-70126 Bari, Italy} 
\maketitle
%%%%%%%%%%%%%%%%%%%%%%%%%%%%%%%%%%%%%%%%%%%%%%%%%%%%%%%%%%%%%%%%%%%
\title{Tomographic test of Bell's inequality}
\author{G. M. D'Ariano, L. Maccone and M. F. Sacchi}
\address{Theoretical Quantum Optics Group\\
Universit\`a degli Studi di Pavia, INFM --- Unit\`a di Pavia, 
via A. Bassi 6, I-27100 Pavia, Italy}
\author{A. Garuccio} \address{Dipartimento Interateneo di Fisica,
INFN, Sezione D, I-70126 Bari, Italy} 
%%%%%%%%%%%%%%%%%%%%%%%%%%%%%%%%%%%%%%%%%%%%%%%%%%%%%%%%%%%%%%%%%%%%%%%
\begin{abstract}
We present a homodyne detection scheme to verify Bell's inequality on
correlated optical beams at the output of a nondegenerate parametric
amplifier. Our approach is based on tomographic measurement of the
joint detection probabilities, which allows high quantum efficiency at
detectors. A self-homodyne scheme is suggested to simplify the
experimental set-up.
\end{abstract}
%%%%%%%%%%%%%%%%%%%%%%%%%%%%%%%%%%%%%%%%%%%%%%%%%%%%%%%%%%%%%%%%%%%%%%%
\section{Introduction}
In 1935 Einstein, Podolsky and Rosen \cite{EPR} proved the
incompatibility among three hypotheses: 1) quantum mechanics is
correct; 2) quantum mechanics is complete; 3) the following criterion
of local reality holds: ``If, without in any way disturbing a system,
we can predict with certainty [...] the value of a physical quantity,
then there exists an element of physical reality corresponding to this
quantity.'' The paper opened a long and as yet unsettled debate about
which one of the three hypotheses should be discarded. While Einstein
suggested to abandon the completeness of quantum mechanics, Bohr
\cite{Bohr} refused the criterion of reality.  The most important step
forward in this debate was Bell's theorem of 1965 \cite{BELL}, which
proved that there is an intrinsic incompatibility between the
assumptions 1) and 3), namely the correctness of quantum mechanics and
Einstein's ``criterion of reality''. In Bell's approach, a source
produces a pair of correlated particles, which travel along opposite
directions and impinge into two detectors. The two detectors measure
two dichotomic observables $A(\alpha)$ and $B(\beta)$ respectively,
$\alpha$ and $\beta$ denoting experimental parameters which can be
varied over different trials, typically the polarization/spin angle of
detection at each apparatus. Assuming that each measurement outcome is
determined by the experimental parameters $\alpha$ and $\beta$ and by
an ``element of reality'' or ``hidden variable'' $\lambda$, Bell
proved an inequality which holds for any theory that satisfies
Einstein's ``criterion of reality'', while it is violated by quantum
mechanics. Such a fundamental inequality, which allows an experimental
discrimination between local hidden--variable theories and quantum
mechanics, has been the focus of interest in a number of experimental
works \cite{Exp}.  \par Unfortunately, Bell's proof is based on two
conditions which are difficult to achieve experimentally. The first is
the feasibility of an experimental configuration yielding perfect
correlation; the second is the possibility of approaching an ideal
measurement, which itself does not add randomness to the
outcome. Since 1969, attention was focused on improving the
correlation of the source on one hand, and, on the other, on deriving
more general inequalities that take into account detection quantum
efficiency or circumvent the problem, however, at the cost of
introducing supplementary hypotheses (see Refs.  \cite{CHSH}), as the
well known ``fair sampling'' assumption. Anyhow it was clear also to
the authors of the same Refs. \cite{CHSH} that these assumptions are
questionable, and, as a matter of fact, it was proved \cite{DcG} that
in all performed experimental checks the results can be reproduced in
the context of Einstein's assumptions if quantum efficiency of
detectors is less than $82.3\%$. However, no experiment has yet
succeeded in realizing such a high value of quantum efficiency.

\par In a typical experiment the source emits a pair of correlated
photons and two detectors separately check the presence of the two
photons after polarizing filters at angles $\alpha $ and $\beta $,
respectively. Alternatively, one can use four photodetectors with
polarizing beam splitters in front, with the advantage of checking
through coincidence counts that photons come in pairs.  Let us denote
by $p_{\alpha ,\beta}$ the joint probability of finding one photon at each 
detector with polarization angle $\alpha $ and $\beta $,
respectively. In terms of the correlation function
\begin{eqnarray}
C(\alpha ,\beta)=&&p_{\alpha ,\beta }+p_{\bar\alpha ,\bar\beta
}-p_{\bar\alpha ,\beta }-p_{\alpha ,\bar\beta }
\label{cab}\;,
\end{eqnarray}  
Bell's inequality \cite{BELL} writes as follows
\begin{eqnarray}B(\alpha ,\beta ,\alpha ',\beta ')\doteq
|C(\alpha ,\beta)-C(\alpha ,\beta ')|+|C(\alpha ',\beta ')+C(\alpha
',\beta)| \leq 2 \;\label{bellineq},
\end{eqnarray}
$\bar\alpha$ and $\bar\beta$ being the polarization angles orthogonal
to $\alpha$ and $\beta$ respectively.  In this letter we propose a new
kind of test for Bell's inequality based on homodyne 
tomography \cite{raymer,breiten} (for a review see
Ref. \cite{bilkent}).  In our set-up the photodetectors are replaced by
homodyne detectors, which along with the tomographic technique can be
regarded as a black box for measuring the joint probabilities
$p_{\alpha ,\beta}$. The main advantage of the tomographic test is
that it allows using linear photodiodes with quantum efficiency $\eta$
higher than $90\%$ \cite{kumar}. On the other hand, the method works
effectively even with $\eta$ as low as $70\%$, without the need of a
``fair sampling'' assumption, since all data are collected in a single
experimental run. With respect to the customary homodyne technique,
which in the present case would need many beam splitters and local
oscillators (LO) that are coherent each other, the set-up is greatly
simplified by using the recent self-homodyne technique
\cite{kumdar}. Another advantage of self-homodyning is the more
efficient signal-LO mode-matching, with improved overall quantum
efficiency.
\section{The experimental set-up}
\begin{figure}[hbt]
\vskip .5truecm
\begin{center}\epsfxsize=.8 \hsize\leavevmode\epsffile{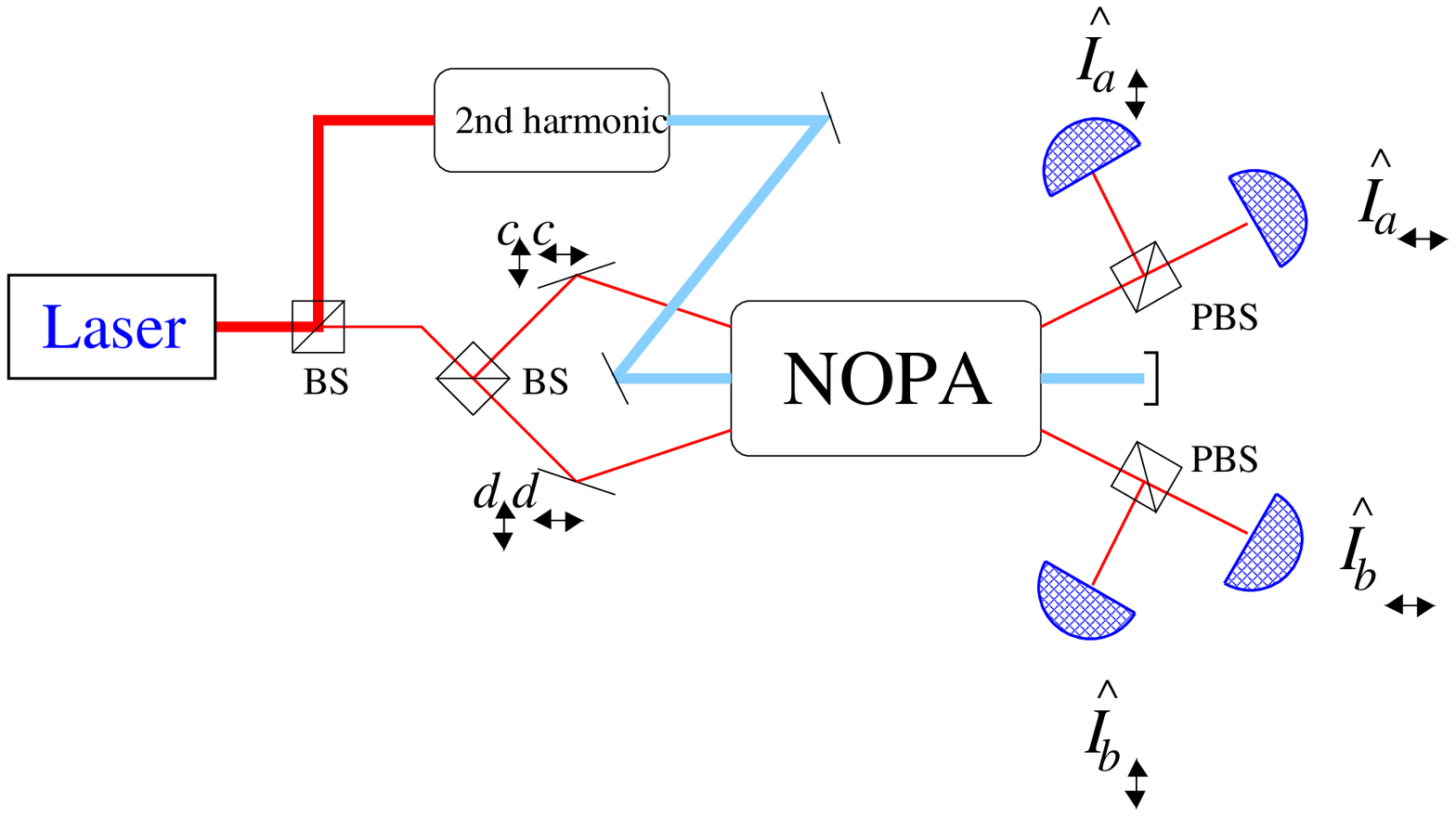}
\end{center}
\caption{Experimental set-up for the tomographic test of Bell's
inequality. PBS and BS denote `polarizing beam splitter' and
`conventional beam splitter' respectively. Input radiation modes
$a_\s$, $b_\h$, $a_\h$ and $b_\s$ are in the vacuum state, while modes
$c_\s$, $c_\h$, $d_\s$, $d_\h$ (at laser frequency $\omega_0$) are in
a coherent state. At the output of the nondegenerate parametric
amplifier (NOPA) the four photocurrents $\hat I$ at radiofrequency
$\Omega$ are measured, yielding the value of quadratures of the field
modes $a_\s$, $b_\h$, $a_\h$ and $b_\s$. The outcome quadratures are
then used to reconstruct the probabilities of interest through quantum
tomography.}
\label{f:experiment}\end{figure}

The apparatus for generating the correlated beams is a $\chi^{(2)}$
nonlinear crystal, cut for Type-II phase--matching, acting as a
nondegenerate optical parametric amplifier (NOPA). The NOPA is
injected with excited coherent states (see Fig. \ref{f:experiment}) in
modes $c_{\h},c_{\s},d_{\h},d_{\s }$ all with equal intensities and at
the same frequency $\omega_0$, $c$ and $d$ denoting mode operators for
the two different wave-vector directions, and $\s$ and $\h$
representing vertical and horizontal polarization, respectively. The
NOPA is pumped at the second harmonic $2\omega_0$. 
At the output of the amplifier four photodetectors
separately measure the intensities ${\hat I}_{a_\s}$, ${\hat
I}_{b_\h}$, ${\hat I}_{a_\h}$, ${\hat I}_{b_\s} $ of the mutual
orthogonal polarization components of the fields propagating at
different wave vectors. A narrow band of the photocurrent is selected,
centered around frequency $\Omega\ll\omega_0$ (typically $\omega_0$ is
optical/infrared, whereas $\Omega$ is a radio frequency).  In the
process of direct detection, the central modes $c_{\s,\h}$ and
$d_{\s,\h}$ beat with $\omega_{0}\pm\Omega$ sidebands, thus playing
the role of the LO of homodyne detectors. The four photocurrents
${\hat I}_{a_\s}$, ${\hat I}_{b_\h}$, ${\hat I}_{a_\h}$, ${\hat
I}_{b_\s} $ yield the value of the quadratures of the four modes
\cite{kumdar}
\begin{eqnarray}
s_{\pi }=\frac{1}{\sqrt{2}} \Big(a_{\pi }(+)+a_{\pi }(-)\Big
)\,,\qquad s=\{a,b\}\,,\qquad \pi=\{\h,\s\}
\label{summodes}\;,
\end{eqnarray}  
where $a_{\pi}(\pm)$ and $b_{\pi}(\pm)$ denote the sideband modes at
frequency $\omega_0\pm\Omega$, which are in the vacuum state at the
input of the NOPA. The quadrature is defined by the operator $\hat
x_\phi\doteq\frac 12(a e^{-i\phi}+a^\dag e^{i\phi})$, where $\phi$ is 
the relative phase between the signal and the LO. The
value of the quadratures is used as input data for the tomographic
measurement of the correlation function $C(\alpha,\beta)$. 
The direction of polarizers $(\alpha,\beta)$ in the experimental 
set-up does not need to be varied over 
different trials, because, as we will show in the following, such direction 
can be changed tomographically.
\par We will now enter into details on the state at the output of the
NOPA and on the tomographic detection. In terms of the field modes in
Eq. (\ref{summodes}) the spontaneous down-conversion at the NOPA is
described by the unitary evolution operator
\begin{equation}
\hat U(\xi )=\exp\left[ \xi\left(a_{\s}^{\dag }b_{\h}^{\dag }+
e^{i\varphi }a_{\h}^{\dag }b_{\s}^{\dag }\right)-\hbox{h. c.}\right]
\label{unit}\;,
\end{equation}  
where $\xi=\chi^{(2)}\gamma L/c$ is a rescaled interaction time
written in terms of the nonlinear susceptibility $\chi ^{(2)}$ of the
medium, the crystal length $L$, the pump amplitude $\gamma$ and the
speed $c$ of light in the medium, whereas $\varphi $ represents a
tunable phase that can be varied by rotating the crystal around the
optical axis \cite{dema}. 
The state at the output of the NOPA writes as follows
\begin{eqnarray}
|\psi\rangle =(1-|\Lambda |^2)\sum _{n=0}^{\infty } \sum
_{m=0}^{\infty } \Lambda ^{n+m} e^{i\varphi m} |n,n,m,m\rangle
\equiv |\psi _{1,2}\rangle \otimes|\psi _{3,4}\rangle
\label{psi}\;,
\end{eqnarray}  
where $\Lambda =\xi/{|\xi|}\; \tanh |\xi|$ and $|i,l,m,n \rangle $
represents the common eigenvector of the number operators of modes
$a_{\s},\ b_{\h},\ a_{\h},\ b_{\s}$, with eigenvalues $i,l,m$ and $n$,
respectively. The average photon number {\em per} mode is given by 
$N=|\Lambda |^2/(1-|\Lambda |^2)$. 
The four-mode state vector in Eq. (\ref{psi})
factorizes into a couple of twin beams $|\psi _{1,2}\rangle $ and
$|\psi _{3,4}\rangle $, each one entangling a couple of spatially
divergent modes ($a_{\s}$, $b_{\h}$ and $a_{\h}$, $b_{\s}$,
respectively).  \par Notice that conventional experiments, concerning
a two-photon polarization-entangled state generated by spontaneous
down-conversion, consider a four-mode entangled state which
corresponds to keeping only the first-order terms of the sums in
Eq. (\ref{psi}), and to ignoring the vacuum component, as only
intensity correlations are usually measured.  Here, on the contrary,
we measure the joint probabilities on the state (\ref{psi}) 
to test Bell's inequality through homodyne tomography, 
which yields the value of $B(\alpha ,\beta ,\alpha ',\beta ')$ in Eq. 
(\ref{bellineq}). 
\section{Tomographic test of Bell's inequality}
The tomographic technique 
is a kind of universal detector, which
can measure any observable $\hat O$ of the field, by averaging a
suitable ``pattern'' function ${\cal R}[\hat O](x,\phi )$ over
homodyne data $x$ at random phase $\phi $. The ``pattern'' function is
obtained through the trace-rule \cite{tokio}
\begin{eqnarray}
{\cal R}[\hat O](x,\phi )= \hbox{Tr}\left[\hat O K_{\eta}(x-\hat
x_{\phi })\right]\;,
\end{eqnarray}
where $K_{\eta }(x)$ is a distribution given in Ref. \cite{DLP}. For
factorized many-mode operators $\hat O=\hat O_1\otimes\hat O_2\otimes
...  \otimes \hat O_n$ the pattern function is just the product of
those corresponding to each single-mode operator $\hat O_1,...,\hat
O_n$ labelled by variables $(x_1,\phi _1),...,(x_n,\phi _n)$.  By
linearity the pattern function is extended to generic many-mode operators.
\par Now we consider which observables are involved in testing Bell's
inequality (\ref{bellineq}). Let us denote by 
$q_{\alpha ,\beta }(i,l,m,n)$ the
probability of having $i,l,m,n$ photons in modes $a_{\s
},b_{\h},a_{\h},b_{\s}$ for the ``rotated'' state
\begin{eqnarray}
|\psi\rangle _{\alpha ,\beta }\equiv \hat U_{1,3}(\alpha)\hat
U_{2,4}(\beta) |\psi\rangle
\label{psir}\;,
\end{eqnarray}  
$\hat U_{1,3}(\alpha)$ and $\hat U_{2,4}(\beta) $ being the unitary
operators
\begin{eqnarray}
&&\hat U_{1,3}(\alpha)=\exp\left[\alpha \left(a_{\s}^{\dag
}a_{\h}-a_{\s}a_{\h}^{\dag }\right)\right]
\label{polar1}\;,
\\ &&\hat U_{2,4}(\beta) =\exp\left[\beta \left(b_{\s}^{\dag
}b_{\h}-b_{\s}b_{\h}^{\dag }\right)\right]
\label{polar2}\;.
\end{eqnarray}
The probabilities in Eq. (\ref{cab}) can be written as 
$p_{\alpha ,\beta }=p_{\alpha ,\beta }(1,1)$,  
$p_{\bar\alpha ,\bar\beta } = p_{\alpha ,\beta }(0,0)$,  
$p_{\bar\alpha ,\beta } = p_{\alpha ,\beta }(0,1)$,  
and $p_{\alpha ,\bar\beta } = p_{\alpha ,\beta }(1,0)$,  
with 
\begin{eqnarray}
p_{\alpha ,\beta }(n,m)=\frac{q_{\alpha ,\beta }(n,1-m,1-n,m)}{P(1,1)}
\label{prob}\;,
\end{eqnarray}  
and $\{n,m=0,1\}$. 
The denominator $P(1,1)$ in Eq. (\ref{prob})
represents the absolute probability of having at the output of the
NOPA one photon in modes $a_{\s },a_{\h}$ and one photon in modes
$b_{\s },b_{\h}$, independently on the polarization, namely
\begin{eqnarray}
P(1,1)={\sum _{n=0,1}\sum _{m=0,1}q_{\alpha ,\beta}(n,1-m,1-n,m)}
\label{punc}\;. 
\end{eqnarray}  
Notice that our procedure does not need a fair sampling assumption,
since we measure in only one run, both the numerator and the
denominator of Eq. (\ref{prob}), namely we do not have to collect
auxiliary data to normalize probabilities.  On the other hand, since
we can exploit quantum efficiencies as high as $\eta =90\%$ or more,
and the tomographic pattern functions already take into account $\eta
$, we do not need supplementary hypothesis for it.  \par The
ob\-ser\-vab\-les that corres\-pon\-d to probabi\-li\-ties
$q_{\alpha,\beta}(i,l,m,n)$ in Eqs. (\ref{prob}) and (\ref{punc}) are
the following 
\begin{eqnarray}
|\,i,l,m,n\rangle _{\alpha ,\beta }\ \ {}_{\alpha ,\beta }\langle
\,i,l,m,n | 
=\hat U^{\dag }_{1,3}(\alpha )\,\hat
U^{\dag }_{2,4}(\beta )\,|\,i,l,m,n\rangle \langle \,i,l,m,n
|\,\hat U_{2,4}(\beta )\,\hat U_{1,3}(\alpha )\;.\label{proj}
\end{eqnarray}
After a straightforward calculation using Eqs. (\ref{prob}),
(\ref{punc}) and (\ref{proj}), one obtains that $P(1,1)$ is measured
through the following average $\cal AV$ of homodyne data
\begin{eqnarray}
P(1,1)={\cal
AV}\left\{\left(K_1^1\,K_0^3+K_0^1\,K_1^3\right)
\left(K_1^2\,K_0^4+K_0^2\,K_1^4\right) \right\}\;,\label {k2}
\end{eqnarray} 
where $K_n^j$ denotes the diagonal $(n=0,1)$ tomographic kernel
function for mode $j$, namely
\begin{eqnarray}
K_n^j \equiv \langle n |K_{\eta}(x-\hat x_{{\phi }_j})|n\rangle
\;.\label{diag}
\end{eqnarray}
The probabilities in the numerator of Eq. (\ref{prob}) are given by the
average of a lengthy expression, which depends on both the diagonal
terms (\ref{diag}) and the following off-diagonal terms
\begin{eqnarray}
K_+^j \equiv \langle 0|K_{\eta}(x-\hat x_{{\phi }_j})|1\rangle\;,
\quad\qquad K_-^j \equiv \langle 1|K_{\eta}(x-\hat x_{{\phi
}_j})|0\rangle = (K_+^j )^*\;.\label{ud}
\end{eqnarray}
Here we report the final expression for $C(\alpha,\beta)$ of
Eq. (\ref{cab})
\begin{eqnarray}
C(\alpha ,\beta)=\frac{1}{P(1,1)}{\cal AV}&&
\Big\{\big[\cos(2\alpha)\left(K_1^1\,K_0^3-K_0^1\,K_1^3\right)
\label {cab2}
+\sin(2\alpha)\left(K_+^1\,K_-^3+K_-^1\,K_+^3\right)\big]\times
\nonumber\\
&& \ \ \big[\cos(2\beta)\left(K_0^2\,K_1^4-K_1^2\,K_0^4\right)
+\sin(2\beta )\left(K_+^2\,K_-^4+K_-^2\,K_+^4  
\right)\big]\Big\}\,.
\end{eqnarray}
\par Caution must be taken in the estimation of the statistical error,
because $C(\alpha,\beta)$---and thus $B(\alpha,\beta,\alpha',\beta')$
in Eq.  (\ref{bellineq})---are non linear averages (they are ratios of
averages). The error is obtained from the variance calculated upon
dividing the set of data into large statistical blocks. However, since
the nonlinearity of $B$ introduces a systematical error which is
vanishingly small for increasingly larger sets of data, the estimated
mean value of $B$ is obtained from the full set of data, instead of
averaging the mean value of blocks.  
\section{Numerical results}
\begin{figure}[hbt]
\vskip .5truecm
\begin{center}\epsfxsize=.4 \hsize\leavevmode\epsffile{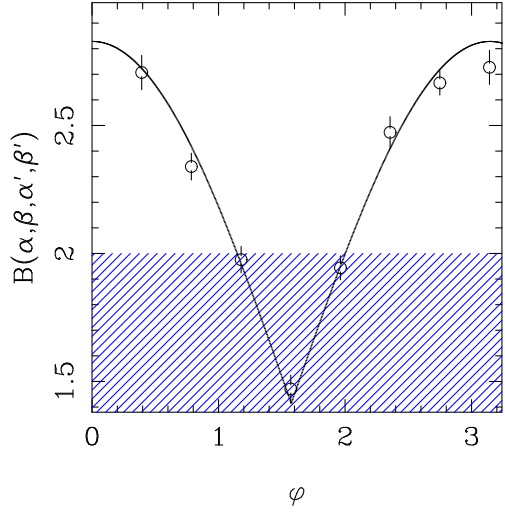}
\end{center}
\caption{Plot of $B(\alpha,\beta,\alpha',\beta')$ {\it vs} the phase
$\varphi$ in the state of Eq. (\ref{psi}) for a simulated
experiment. The shaded area represents the classical region for
$B$. The parameters of the simulation are: $\alpha=0;\ \beta=\frac
38\pi;\ \alpha'=\frac \pi4;\ \beta'=\frac \pi8$; quantum efficiency
$\eta=85\%$; average photon number {\em per} mode $N=0.5$. A total number of
$10^6$ homodyne data (divided into $20$ statistical blocks) has been
used.}
\label{f:ciclophi}\end{figure}

\begin{figure}[hbt]
\vskip .5truecm
\begin{center}\epsfxsize=.4 \hsize\leavevmode\epsffile{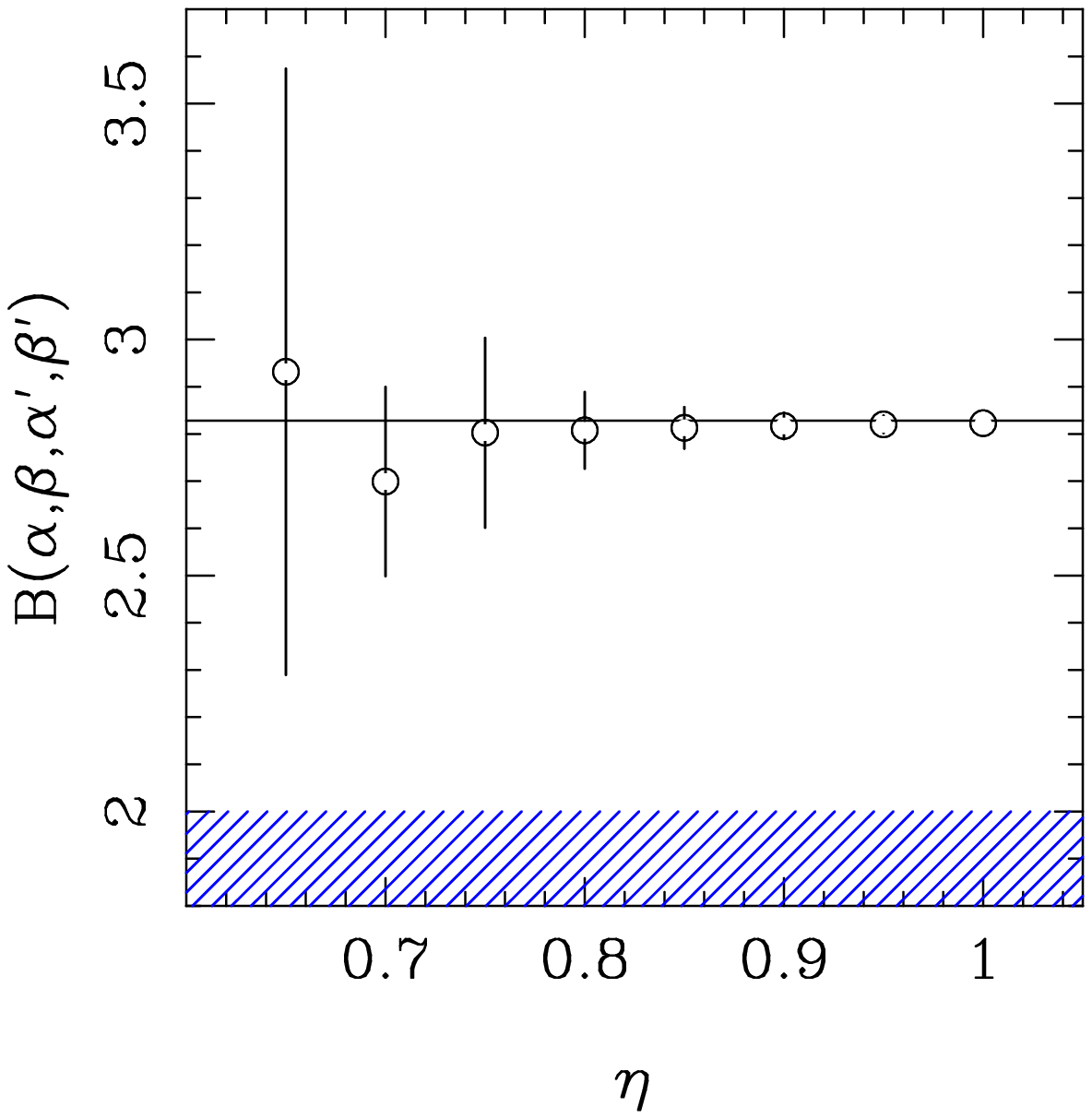}
\end{center}
\caption{Plot of $B(\alpha,\beta,\alpha',\beta')$ {\it vs} the quantum
efficiency of the detectors for a series of simulated experiments. The
shaded area represents the classical region for $B$. The parameters of
the simulations are: $\alpha=0;\ \beta=\frac 38\pi;\ \alpha'=\frac
\pi4;\ \beta'=\frac \pi8$; $\varphi=\pi$; $N=0.5$. A total number
of $6\cdot 10^7$ homodyne data (in 20 statistical blocks)
has been used for each simulation.}
\label{f:cicloeta}\end{figure}

We now present some numerical
results obtained from Monte--Carlo simulations of the proposed
experiment. For the simulation we use the theoretical homodyne probability
pertaining to the state $|\psi\rangle$ in Eq. (\ref{psi}) which, for
each factor $|\psi _{i,j}\rangle $, is given by
\begin{eqnarray}
p_{\eta }(x_i,x_j;\phi _i,\phi _j)\!=\! {2\exp\left[
-{{(x_i+x_j)^2}\over{d^2_{z_{ij}}+4\Delta^2_{\eta}}}-
{{(x_i-x_j)^2}\over{d^2_{-z_{ij}}+4\Delta^2_{\eta}}}\right]
\over{\pi\sqrt{(d^2_{z_{ij}}+4\Delta^2_{\eta})(d^2_{-z_{ij}}+4
\Delta^2_{\eta})}}}\,, \label{pxy}
\end{eqnarray}
where $x_i$ ($i=1,2,3,4$) is the outcome of the homodyne measurement
for quadrature of the $i$-th mode at phase $\phi_i$, and
\begin{eqnarray}
z_{ij}=e^{-i(\phi _i+\phi _j)}\Lambda\;,\qquad d^2_{\pm
z_{ij}}={{|1\pm z_{ij}|^2}\over{1-|z_{ij}|^2}}\;,\qquad\Delta
^2_{\eta }=\frac{1-\eta}{4\eta}\;.
\end{eqnarray} 
In Fig. \ref{f:ciclophi} we present the simulation results for $B$ in
Eq.  (\ref{bellineq}) {\em vs} the phase $\varphi$ in the state of
Eq. (\ref{psi}).  The full line represents the value of $B$ in
Eq. (\ref{bellineq}) with the quantum theoretical value $C(\alpha
,\beta)$ given by
\begin{eqnarray}
C(\alpha ,\beta )=\cos\varphi\sin 2\alpha \sin 2\beta -\cos
2\alpha\cos 2\beta
\label{cabteor}\;.
\end{eqnarray}  
Quantum efficiency $\eta=85\%$ has been used, nonetheless notice that
for $\varphi=\pi$ (corresponding to a maximum violation with respect
to the classical bound $2$), the obtained value is over $10\;\sigma$
distant from the bound. By increasing the number of homodyne data, it
is possible to obtain good results also for lower quantum
efficiency. In fact, by increasing the number of data to $8\cdot
10^8$, a value of $B(0,\frac 38\pi,\frac\pi 4,\frac\pi 8)=2.834\pm
0.268$ has been obtained for $\Lambda=0.5$, $\varphi=\pi$, and $\eta$
as low as $65\%$.  This result is to be compared with the quantum
theoretical value of $2\sqrt{2}$. In Fig. \ref{f:cicloeta} the results
of the measurement of $B$, for different simulated experiments using
the same number of data, are presented for different detector
efficiencies $\eta$. Notice how the error bars decrease {\em versus}
$\eta $.  \par For an order of magnitude of the data acquisition rate
in a real experiment, one can consider that in a typical self-homodyne
set-up with a NOPA pumped by a $2^{\scriptsize nd}$ harmonic of a
Q-switched mode-locked Nd:YAG laser, the Q-switch and the mode-locker 
repetition rates are 10 kHz and 100 MHz, respectively. 
Typical time of the boxcar
integration is 10 ns, so that one sample per pulse can be
collected. In summary, $10^7$ data samples can be obtained in $10^3$
s.  In such an experimental arrangement, for a Q-switch envelope of
200 ns, the shot noise can be reached by the peak amplitude of the
5-MHz low-pass-filtered photocurrent. For more detailed experimental
parameters, the reader is referred to Ref. \cite{opt}. 
\section{Conclusions}
In conclusion we have proposed a test of Bell's inequality, based
on self--homodyne tomography. The rather simple experimental apparatus
is mainly composed of a NOPA crystal and four photodiodes. The
experimental data are collected through a self--homodyne scheme and
processed by the tomographic technique. No supplementary hypotheses are
introduced, a quantum efficiency $\eta $ as high as $90\%$ is
currently available, and, anyway, $\eta $ as low as $70\%$ is
tolerated for $10^6$--$10^7$ experimental data. We have presented
some numerical results based on Monte--Carlo simulations that confirm
the feasibility of the experiment, showing violations of Bell's
inequality for over $10\;\sigma$ with detector quantum efficiency 
$\eta=85\%$.
\section*{Acknowledgments}
The authors thank the anonymous referee for his/her useful suggestions.  
The Quantum Optics Group of Pavia acknowledges the INFM for
financial support (PRA--CAT97).

%%%%%%%%%%%%%%%%%%%%%%%%%%%%%%%%%%%%%%%%%%%%%%%%%%%%%%%%%%%%%%%%%%%%%
\newpage
%%%%%%%%%%%%%%%%%%%%%%%%%%%%%%%%%%%%%%%%%%%%%%%%%%%%%%%%%%%%%%%%%%%%%
\newpage
%%%%%%%%%%%%%%%%%%%%%%%%%%%%%%%%%%%%%%%%%%%%%%%%%%%%%%%%%%%%%%%%%%%%%
\newpage
%%%%%%%%%%%%%%%%%%%%%%%%%%%%%%%%%%%%%%%%%%%%%%%%%%%%%%%%%%%%%%%%%%%%%

\begin{thebibliography}{99}
\bibitem{EPR} A. Einstein, B. Podolsky, and N. Rosen, Phys. Rev.  {\bf
47}, 777 (1935).
\bibitem{Bohr} N. Bohr, Phys. Rev.  {\bf 48}, 696 (1935). 
\bibitem{BELL} J. S. Bell, Physics  {\bf 1}, 195 (1965).
\bibitem{Exp} J. F. Clauser, Phys. Rev. Lett.  {\bf 36}, 1223
(1976); A. Aspect, J. Dalibard, and G. Roger, Phys. Rev. Lett.  {\bf
47}, 460 (1981); A. J. Duncan, W. Perrie, H. J. Beyer, and
H. Kleinpoppen, in {\em Fundamental Processes in Atomic Collision
Physics}, Plenum, New York, 1985, pg. 555; Z. Y. Ou and L. Mandel,
Phys. Rev. Lett. {\bf 61}, 50 (1988); C. O. Alley and Y. H. Shih,
Phys. Rev. Lett. {\bf 61}, 2921 (1988); J. D. Franson,
Phys. Rev. Lett. {\bf 62}, 2200 (1989); K. Mattle, H. Weinfurter,
P. G. Kwiat, and A. Zeilinger, Phys. Rev. Lett. {\bf 76}, 4656 (1996).
\bibitem{CHSH} J. F. Clauser, M. A. Horne, A. Shimony, and R. A. Holt,
Phys. Rev. Lett. {\bf 23}, 880 (1969); J. F. Clauser and M. A. Horne,
Phys. Rev. D {\bf 10}, 256 (1974); A. Garuccio and V. A. Rapisarda,
Nuov. Cim. A {\bf 65}, 269 (1981).
\bibitem{DcG} L. De Caro and A. Garuccio, Phys. Rev. A {\bf 54}, 174
(1996). 
\bibitem{raymer} D. T. Smithey, M. Beck, M. G. Raymer, and A. Faridani,
Phys. Rev. Lett. {\bf 70}, 1244 (1993). 
\bibitem{breiten} G. Breitenbach, S. Schiller, and J. Mlynek, 
Nature {\bf 387}, 471 (1997).
\bibitem{bilkent} G. M. D'Ariano, {\it Measuring quantum states}, in {\it 
Quantum Optics and the Spectroscopy of Solids}, ed. by T. Hakio\v{g}lu
and A.S. Shumovsky, Kluwer Academic Publishers (1997), p. 175.
\bibitem{kumar} C. Kim and P. Kumar, Phys. Rev. Lett. {\bf 73}, 1605
(1994). 
\bibitem{kumdar} G. M. D'Ariano, M. Vasilyev, and P. Kumar, 
Phys. Rev. A {\bf 58}, 636 (1998).
\bibitem{dema} D.Boschi, F. De Martini, and G. Di Giuseppe, 
in {\em Quantum Interferometry}, F. De Martini, G. Denardo
and Y. Shih, Eds. (VCH, Wenheim 1996), p. 135.
\bibitem{tokio} G. M. D'Ariano, in {\it Quantum Communi\-ca\-tion, 
Com\-pu\-ting, and Mea\-su\-re\-ment}, 
ed. by O. Hirota, A. S. Holevo and C. M. Caves, Plenum 
Publishing (New York and London 1997), p. 253. 
\bibitem{DLP} G. M. D'Ariano, U. Leonhardt, and H. Paul, Phys. Rev. A
{\bf 52}, R1801 (1995). 
\bibitem{opt} M. Vasilyev, S-K Choi, P. Kumar, and G. M. D'Ariano, 
Opt. Lett. {\bf 23}, 1393 (1998).
\end{thebibliography}
\end{document}